 \documentclass[preprint2]{aastex}

\usepackage{natbib}
\usepackage{graphics}
\usepackage{graphicx}

\usepackage{amsmath}
\newcommand{\beq}{\begin{equation}}
\newcommand{\enq}{\end{equation}}
\newcommand{\beqa}{\begin{eqnarray}}
\newcommand{\enqa}{\end{eqnarray}}
\newcommand{\beit}{\begin{itemize}}
\newcommand{\enit}{\end{itemize}}
\newcommand{\bem}{\begin{pmatrix}}
\newcommand{\enm}{\end{pmatrix}}

\newcommand{\Tr}{\mathrm{Tr}}

\newcommand{\lat}{\left\langle}
\newcommand{\rat}{\right\rangle}
\newcommand{\av}[1]{\lat #1 \rat}

\newcommand{\binq}[3]{\bem #1 \\ #2 \enm_{#3}}
\newcommand{\pochh}[3]{\lp #1 : #2 \rp_{#3}}

\newcommand{\lb}{\left [}
\newcommand{\rb}{\right ]}
\newcommand{\lp}{\left (}
\newcommand{\rp}{\right )}

\newcommand{\ca}[1]{s_{#1}}

\renewcommand{\bem}{\begin{bmatrix}}
\renewcommand{\enm}{\end{bmatrix}}
\newcommand{\mlnrho}{\bar{\ln \rho}}
\newcommand{\mrho}{\bar{\rho}}

\begin{document}


\title{On the incompleteness of the moment and correlation function hierarchy as probes of the lognormal field.}


\author{Julien Carron}
\affil{Institute for Astronomy, ETHZ, CH-8093 Zurich, Switzerland}

\email{jcarron@phys.ethz.ch}

\begin{abstract}
We trace with analytical methods and in a model parameter independent manner the independent bits of Fisher information of each of the moments of the lognormal distribution, as a now standard prescription for the distribution of the cosmological matter density field, as it departs from Gaussian initial conditions. We show that, when entering the regime of large fluctuations, only a tiny, dramatically decaying fraction of the total information content remains accessible through the extraction of the full series of moments of the field. This is due to a known peculiarity of highly tailed distributions, that they cannot be uniquely recovered given the values of all of their moments. This renders under this lognormal assumption cosmological probes such as the correlation function hierarchy or equivalently their Fourier transforms fundamentally limited once the field becomes non linear, for any parameter of interest. We show that the fraction of the information accessible from two-point correlations decays to zero following the inverse squared variance of the field. We discuss what general properties of a random field's probability density function are making the correlation  function hierarchy an efficient or inefficient, complete or incomplete set of probes of any model parameter.

\end{abstract}




\section{Introduction \label{introduction}}
The cosmological matter density field is becoming more and more directly accessible to observations with the help of weak lensing  \citep{1992grle.book.....S,2001PhR...340..291B,2003ARA&A..41..645R,2006astro.ph.12667M}. Its statistical properties are the key element in trying to optimize future large galaxy surveys aimed at answering actual fundamental cosmological questions, such as the nature of the dark components of the universe \citep{2009Natur.458..587C,2008ARA&A..46..385F}. To this aim, Fisher's measure of information on parameters \citep{fisher25,Rao,vandenbos07} has naturally become of standard use in cosmology. It provides indeed an handy framework, in which it is possible to evaluate in a quantitative manner the statistical power of some experiment configuration aimed at some observable \citep[e.g.]{1997ApJ...480...22T,Tegmark97,1999ApJ...514L..65H,2004PhRvD..70d3009H,2007MNRAS.381.1018A,2007MNRAS.377..185P,2006astro.ph..9591A,2009ApJ...695..652B}. Such studies are in the vast majority of cases limited to Gaussian probability density functions, or perturbations therefrom, and deal mostly with the prominent members of the correlation function hierarchy \citep{1980lssu.book.....P}, or equivalently their Fourier transforms the polyspectra, such as the matter power spectrum.
\newline\noindent
The approach via the correlation function hierarchy is very sensible in the nearly linear regime for at least two reasons. First, in principle, the correlations are the very elements that cosmological perturbation theory is able to predict in a systematic manner (see \citep{2002PhR...367....1B} for a review, or the more recent \citep{2011PhRvD..83h3518M} and the numerous references in it). Second, primordial cosmological fluctuations fields are believed to be accurately described by the use of Gaussian statistics. It is well known that the correlations at the two-point level provide a complete description of Gaussian fields. It is therefore natural to expect this approach to be adequate throughout the linear and the mildly non linear regime, when departures from Gaussianity are small.
\newline\noindent
Deeper in the non linear regime, fluctuations grow substantially in size, and tails in the matter probability density function do form. A standard prescription for the statistics of the matter field in these conditions is the lognormal distribution, various properties of which are discussed in details in an astrophysical context in \citep{1991MNRAS.248....1C}. It was later shown to be reproduced accurately, both from the observational point of view as well as in comparison to standard perturbation theory and N-body simulations \citep{1995ApJ...443..479B,1994A&A...291..697B,2001ApJ...561...22K,2000MNRAS.314...92T,2005MNRAS.356..247W}, in low dimensional settings. More recently, it was used as a starting point for a tentative of a better description of clustering \citep{2010arXiv1012.3168K}.  The lognormal assumption is also very much compatible with numerical works \citep{2009ApJ...698L..90N,2011ApJ...731..116N} showing that the spectrum of logarithm of the field $\ln 1 + \delta$ carries much more  information than the spectrum of $\delta$ itself. The first evaluation of the former within the framework of perturbation theory appeared recently \citep{2011arXiv1103.2166W}.
\newline\noindent
Lognormal statistics \citep[for a textbook presentation]{Aitchison57} are not innocuous. More specifically, the lognormal distribution is only one among many distributions that leads to the very same series of moments. This fact indicates that, going from the distribution to the moments, one may be losing information in some way or another. A fundamental limitation of the correlation function hierarchy in extracting the information content of the field in the non linear regime could therefore exist, if its statistics are indeed similar to the lognormal. This important fact was already mentioned qualitatively in \citep{1991MNRAS.248....1C}, but it seems no quantitative analysis is available at present.
\newline\noindent
 It is the purpose of this paper to provide first answers to these issues, in terms of Fisher information, looking at the details of the structure of the information within the lognormal field. It is built out of two main parts. \newline\newline
The first deals exclusively with the case of a single lognormal variable, illustrating the main aspects we want to point out in this work. We begin by presenting how to identify the independent bits of information that are contained in the successive moments of a distribution, with the help of orthogonal polynomials. We discuss the properties of this decomposition that are relevant for our purposes. The procedure is very similar to the decompositions presented in \citep{Jarrett84}, to which we refer for a more complete discussion on the properties of such expansions. In a second step, we perform this decomposition for the lognormal distribution, which can be obtained exactly at all order in terms of $q$-series, due to its convenient analytical properties.  It contains one of our main results, presented in figure \ref{fig1} : when the variance of the fluctuations of the lognormal distribution reaches unity, essentially the entirety of its information \footnote{Throughout this work, by 'information' is meant  more rigorously Fisher information.} content cannot be accessed anymore through a study of its moments, even if  the complete series of moments could be extracted. We then delve a little bit more into the details of that phenomenon, and show that this is due to the inability of polynomials to reproduce the logarithm function, leading to missing bits of information in the part of the distribution describing the underdense regions. Finally, we perform a comparison of our results for the lower order moments with results \citep{1994A&A...291..697B} from standard perturbation theory, and find good agreement over several orders of magnitude in the variance.
\newline\newline
The second part extends the analysis to the multivariate lognormal distribution, and to the continuous limit of the lognormal field. The decomposition of the information into uncorrelated pieces is conceptually identical. Due to the highly increased formal complexity, the explicit expressions for the information content of the $n$-points correlations are however of less practical use. We therefore focus on two simpler situations that can be dealt with analytically, dealing with the extraction of the mean of a lognormal field and two point correlations.\newline \newline
We then summarise the results, and conclude with a discussion on the conditions for the correlation function hierarchy to form a complete or incomplete, efficient or inefficient set of probes of a random field. The appendix contains a short series of technical details regarding the material presented in the main text.
\newline\indent
\section{One variable}
For a variable $X$, with probability density function $p(x,\alpha)$, $\alpha$ any parameter, we are assuming that all moments exist, write for them
\beq
m_n =  \av{x^n},\quad n  = 0,1,\cdots
\enq
and the associated covariance matrix with
\beq
\Sigma _{ij} = m_{i+j}- m_im_j
\enq
Since $p(x,\alpha)$ is normalised to unity for any value of the parameter, we have
\beq
\frac{\partial m_0}{\partial \alpha} =  0 = \av{s(x,\alpha)},
\enq
where $s(x,\alpha)  = \partial_\alpha \ln p(x,\alpha) $ is the score function.
Fisher's measure of information \citep{fisher25,vandenbos07} on the parameter $\alpha$ is then defined as the variance of the score function
\beq
F_\alpha = \av{ s^2(x,\alpha)}.
\enq
The Fisher information density
\beq \label{infdensity}
s^2(x,\alpha)p(x,\alpha)dx = \frac{\lp \partial_\alpha p(x,\alpha) \rp^2}{p(x,\alpha)}dx
\enq
is the amount of information associated to observations of realizations of the variable in the range $\lp x,x+dx \rp$. $F_\alpha$ itself has gotten through the Cram\'er-Rao inequality \citep{Rao} the widespread interpretation in cosmology of approximating the error bars the experiment under consideration will be able to put on the parameter $\alpha$ \citep[e.g.]{1997ApJ...480...22T}.
\subsection{Fisher information and orthogonal polynomials \label{s1}}
The decomposition of the information in independent pieces associated to each moment relies on the approximation of the score function through orthonormal polynomials. For each natural number $n$, $P_n$ is a polynomial of degree $n$ and
\beq \label{ortho}
\av{P_m(x)P_n(x)} = \delta_{mn}.
\enq
These polynomials  can always be constructed for a given distribution and are unique up to a sign, which is fixed by requiring the coefficient of $x^n$ in $P_n$ to be positive.  We refer to the textbooks \citep{Szego39,Freud71} for the general theory of orthogonal polynomials. These polynomials can be written in the monomial basis with the help of a triangular transition matrix $C$ that we will use later on,
\begin{equation} \label{cmatrix}
P_n(x)= \sum_{m = 0}^n\:C_{nm}\:x^m.
\end{equation}
According to equation (\ref{ortho}), it holds that the non constant orthogonal polynomials average to zero. As the value of the model parameter changes,  these averages take non vanishing values, at a rate which is equal to the component of the score function parallel to these polynomials :
\beq \label{scoeff}
\frac{\partial \av{P_n(x)}}{\partial \alpha}  = \av{s(x,\alpha) P_n(x)} =: \ca n,
\enq
where the relation $\partial_\alpha p(x,\alpha) = s(x,\alpha)p(x,\alpha)$ was used.
We argue that $s_n^2$ is precisely the independent information content of the moment of order $n$. This can be seen as the following. For any natural number $n$, it is not difficult to show that the inverse covariance matrix of size $n$ is given by
\beq
\lb \Sigma^{-1}\rb_{ij} = \lb C^TC \rb_{ij},\quad i,j = 1,\cdots,n.
\enq
Therefore, noting that from equation (\ref{cmatrix}) and from the definition (\ref{scoeff}) of the information coefficients $s_n$ we can write
\beq \label{scoeff2}
s_n = \sum_{k = 1}^nC_{nk}\frac{\partial m_k}{\partial \alpha},
\enq
the following relation holds,
\beq \label{infn}
\sum_{i = 1}^n s_i^2 =  \sum_{i,j = 1}^n\frac{\partial m_i}{\partial \alpha}\lb \Sigma^{-1}\rb_{ij}\frac{\partial m_j}{\partial\alpha}.
\enq
This expression, weighting the sensitivity of the moments to the parameter by the covariance matrix, is the amount of information present in the first $n$ moments, taking all correlations into account. For instance, this is exactly the amount of information available, if the first $n$ moments were to be extracted, with the help of unbiased, Gaussian distributed estimators \citep[e.g.]{1997ApJ...480...22T}. Also, it is simple to show that these coefficients are invariant under any linear transformation of the field. This allow us to identify unambiguously the $n$th squared coefficient as the independent bits of information contained in the moment of order $n$.
\newline\newline
If the set of orthonormal polynomials forms a complete basis, the partial sums
\beq \label{series}
\sum_{n = 1}^N s_nP_n(x)
\enq
will tend, with increasing $N$, to reproduce accurately the score function. By Parseval identity, the full amount of Fisher information can be written as
\beq
F_\alpha = \sum_{n = 1}^\infty s_n^2.
\enq
This last equation implies that the information contained in the full set of moments is identical to the total amount of information.
This is certainly in perfect agreement with expectations. A well known result due to M. Riesz \citep{Riesz23} in the theory of moments states namely that if the moment problem associated to the moments $\left \{m_i \right\}_{i = 0}^\infty$ is determinate, (i.e. the distribution giving rise to these moments is uniquely determined by their values), then the set of associated orthonormal polynomials is complete. Since the distribution is uniquely determined, common sense would require then the total amount of information contained in the distribution to be the same as the one contained in the full set of moments.
\newline\newline
However, moment problems are not always determinate, and orthonormal polynomials associated to weight functions do not always form complete sets. Therefore, the series
\beq
f_\alpha := \sum_{n = 1}^\infty s_n^2
\enq
may not always converge to the total amount of Fisher information, and, if not, will always underestimate it. We have namely, instead of Parseval's identity, the Bessel inequality,
\beq \label{MSE} \begin{split}
0 & \le \av{\: \lp \sum_{n = 1}^\infty \ca n P_n(x) - s(x,\alpha) \rp^2} \\
&= F_{\alpha} - f_\alpha.
\end{split}
\enq
In words, the mean squared error in approximating the score function with polynomials is the amount of Fisher information absent from the full set of moments. As emphasized already in a astrophysical context by \citep{1991MNRAS.248....1C} and stated in our introduction the moments of the lognormal distribution are precisely an example of an indeterminate moment problem. In fact, a whole family of distribution, given explicitly in \citep{Heyde63}, do have the very same series of moments. In light of these considerations, our subsequent results cannot be considered as surprising.
\newline Before turning to the actual calculation of the coefficients $s_n$ of the lognormal distribution, let us just state that when the score function is itself a polynomial, it is clear that the series (\ref{series}) actually terminates,
\beq
s_k = 0 ,\quad k > n  
\enq
where $n$ is the order of the polynomial representing the score function. The prime example being the Gaussian distribution, for which $n = 2$, with associated orthonormal polynomials the Hermite polynomials. The well known fact that the mean and the variance of the Gaussian distribution carry all of the information becomes within our framework that only the coefficients $s_1$ and $s_2$ are non-zero. 
\subsection{Basic properties of the lognormal distribution}
The variable $X$ has a lognormal distribution when $Y = \ln X$ is a normal variable, with mean $\mu_Y$ and variance $\sigma^2_Y$. The dependency on a model parameter $\alpha$ can enter one or both of these parameters.
The moments of $X$ are given by Gaussian integrals and read explicitly
\beq \label{moments}
m_n =  \exp \lp n\mu_Y + \frac 12 n^2\sigma^2_Y\rp.
\enq
The mean and variance of $Y$ relate therefore to the mean $\mu$ and variance $\sigma^2$ of $X$ according to
\beq \label{rel}
\begin{split}
\sigma^2_Y &= \ln \lp 1 + \sigma^2_\delta \rp \\
\mu_Y &= \ln \mu - \frac 12 \ln \lp 1 + \sigma^2_\delta \rp,
\end{split}
\enq
where $\sigma^2_\delta$ is the variance of the fluctuations of $X$,
\beq
\sigma_\delta^2 = \frac{\sigma^2}{\mu^2}.
\enq
Since $Y$ is has a Gaussian distribution and Fisher's measure is invariant under invertible transformations, the total Fisher information content of $X$ is given by the well known expression for the Gaussian distribution,
\begin{equation} \label{exactlnnormal}
F^X_{\alpha} = \frac{1}{\sigma^2_Y}\lp \frac{\partial \mu_Y}{\partial \alpha} \rp^2 + \frac 1 {2\sigma^4_Y} \lp \frac{\partial \sigma^2_Y}{\partial \alpha} \rp^2.
\end{equation}
The key parameter throughout this part of this work  will be the quantity $q$, defined as
\beq
q := e^{-\sigma^2_Y} = \frac{1}{1 + \sigma^2_\delta}. 
\enq
Note that $q$ is strictly positive and smaller than unity.  The regime of small fluctuations, where the lognormal distribution is very close to the Gaussian distribution is described by values of $q$ close to unity. Deep in the non linear regime, it tends to zero. These two regimes are conveniently separated at $q = 1/2$, corresponding to fluctuations of unit variance. We note the following convenient property of the moments for further reference,
\beq \label{separability}
m_{i + j} = m_im_j \:q^{-ij}.
\enq
\subsection{Information coefficients}
From equations (\ref{scoeff2}), (\ref{moments}), and (\ref{rel}), we see that the $n$-th information coefficient $s_n$ is given by
\beq \label{infcoeff} \begin{split}
s_n &= \frac{\partial \ln \mu}{\partial \alpha} \sum_{k = 0}^n C_{nk}\:m_k \:k \\
&+ \frac 1{2\lp 1 + \sigma^2_\delta\rp}\frac{\partial \sigma^2_\delta}{\partial \alpha} \sum_{k = 0}^n C_{nk}\:m_k \:k(k-1). 
\end{split}
\enq
Evaluation of the above sums can proceed in different ways. Notably, it is possible to get an explicit formula for the orthonormal polynomials, and therefore of the matrix $C$, for the lognormal distribution. These are essentially the Stieltjes-Wigert polynomials \citep{Wigert23,Szego39}. We will namely use their specific form later in this work, though they are not needed for the purpose of evaluating (\ref{infcoeff}). We proceed with the following trick : we introduce the $q$-shifted factorial , also called $q$-Pochammer symbol \citep[ section 10]{Kac02,Andrews99}, as
\beq
\pochh t q n = \prod_{k = 0}^{n-1}\lp 1 - tq^k\rp,\quad \pochh t q 0  := 1
\enq
$t$ a real number, and prove in the appendix that the following curious identity holds,
\beq \label{lemma}
\av{P_n(tx)} =(-1)^n\frac{q^{n/2}}{\sqrt{ \pochh qqn }} \pochh t q n.
\enq
By virtue of
\beq
\av{P_n(tx)} = \sum_{k = 0}^nC_{nk}\:m_k\:t^k.
\enq
it follows from our identity (\ref{lemma}) that the sums given in the right hand side of equation (\ref{infcoeff}) are proportional to the first, respectively the second derivative of the $q-$Pochammer symbol evaluated at $t = 1$. Besides, matching the powers of $t$ on both sides of equation (\ref{lemma}) will provide us immediately the explicit expression for the matrix elements $C_{nk}$.
\newline
\newline\newline
We distinguish explicitly two situations, labelled by an index $a$ taking values $\mu$ or $\sigma$, where only one of the two parameters of the lognormal distribution actually depends on $\alpha$. The general case being reconstructed trivially from these two. 
\paragraph{case $a = \mu$} We assume in this case that the parameter enters the mean of the distribution only,
\beq
\frac{\partial \sigma^2_\delta}{\partial \alpha} = 0
\enq
From (\ref{infcoeff}), we see that the derivative of  $\mu$ with respect to $\alpha$ only plays the role of an overall normalization constant. Since we will deal exclusively with ratios, it is irrelevant for our purposes.  We choose for convenience
\beq
\quad \frac{\partial \ln \mu}{\partial \alpha} = 1.
\enq
The total amount of information in the distribution becomes, from (\ref{exactlnnormal}) and (\ref{rel}),
\beq \label{Fmuex}
F_\alpha^\mu := \frac{1}{\ln \lp  1 + \sigma^2_\delta\rp}.
\enq
\paragraph{case $a = \sigma$}The parameter enters the variance of the distribution only, and we pick again a convenient normalization of its derivative
\beq
\frac 1{2\lp 1 + \sigma^2_\delta\rp} \frac{\partial \sigma^2_\delta}{\partial \alpha} = 1,\quad \frac{\partial \ln \mu}{\partial \alpha} = 0.
\enq
This situation is the most common in cosmology, for instance for any model parameter entering the matter power spectrum.
The exact amount of information becomes, again from (\ref{exactlnnormal}) and (\ref{rel}),
\beq \label{Fsigex}
F_\alpha^\sigma := \frac 1 {\ln \lp  1 + \sigma^2_\delta\rp}\lp 1  + \frac 2 {\ln\lp 1 + \sigma^2_\delta\rp} \rp.
\enq
\newline\newline
In both of these situations, we obtain the information coefficients (\ref{infcoeff}) by differentiating once, respectively twice, our relation (\ref{lemma}) with respect to the parameter $t$, and evaluating these derivatives at $t = 1$. The result is, in the $a = \mu$ case,
\begin{equation} \label{anbn}
\begin{split}
 s_n^\mu  = (-1)^{n-1}\sqrt{ \frac{q^n}{1 - q^n}\pochh q q {n-1} }
 \end{split}
 \end{equation}
 and for $a = \sigma$,
 \begin{equation}\label{Ansigma}
 \begin{split}
 s_n^\sigma  =  - 2s_n^\mu \lb \sum_{k = 1}^ {n-1} \frac{q^k}{1 - q^k}\rb ,\quad n > 1 
\end{split}
\end{equation}
whereas $s_{n = 1}^\sigma$ is easily seen to vanish from its definition.
\newline\newline
\subsection{Incompleteness of the information in the moments}
\begin{figure}
  \includegraphics[width = 0.5\textwidth]{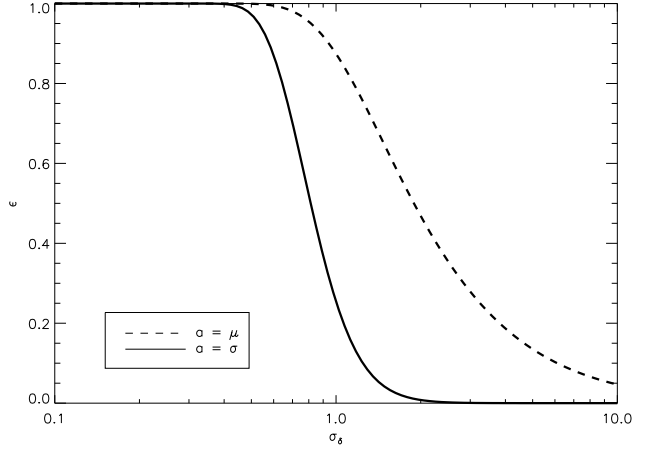}
 \caption{\label{fig1} The fraction of the total information content that is accessible through extraction of the full series of moments of the lognormal field, as function of the square root of the variance of the fluctuations.}
 \end{figure}

\begin{figure}
  \includegraphics[width = 0.5\textwidth]{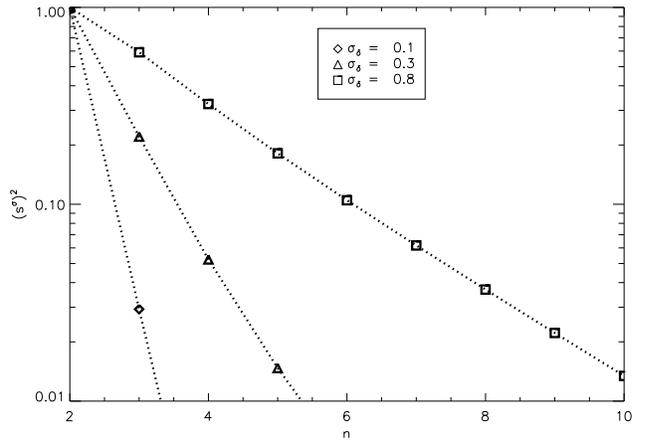}
 \caption{\label{fig2} The distribution of the information within the first 10 moments of the lognormal field, given by the coefficients $\lp s^\sigma_n \rp^2$, equation (\ref{Ansigma}), normalized to the information content of the second moment, for three different values of $\sigma_\delta$. Note that deeper in the non linear regime, the trend is reversed.}
 \end{figure}

The series
\beq \label{fmoments}
 f_\alpha^a =  \sum_{n =1}^\infty \lp s_n^a\rp^2 ,\quad a = \mu,\sigma
\enq
are the total amount of information contained in the full series of moments, in the respective cases described above. The ratios $\epsilon_\mu$ and $\epsilon_\sigma$, defined as
\beq
\epsilon_a := \frac{f_\alpha^a} {F^a_\alpha} ,\quad a = \mu,\sigma.
\enq
are the fraction of the information that can be accessed by extraction of the full set of moments of $X$. The two asymptotic regimes of very small and very large fluctuation variance $\sigma_\delta$ can be seen without difficulty. In both cases, it is seen that the first non vanishing term of the corresponding series dominates completely its value. For very small variance, or equivalently $q$ very close to unity, $\epsilon_a$ tends to unity, illustrating the fact the distribution becomes arbitrary close to Gaussian : all the information is contained in the first two moments. The large variance regime is more interesting, and, even tough the information coefficients decays very sharply as well, the series (\ref{fmoments}) are far from converging to the corresponding expressions (\ref{Fmuex}) and (\ref{Fsigex}) showing the total amount of information. Considering only the dominant first term in the relevant series and setting $q \rightarrow 0$, one obtains 
\begin{equation} \label{asymptotics}
\epsilon_\mu \rightarrow \frac{1}{\sigma_\delta^2} \ln  \lp 1 + \sigma_\delta^2\rp .
\end{equation}
and a much more dramatic decay of $\epsilon_\sigma$ :
\begin{equation} \label{asymptotics2}
\epsilon_\sigma \rightarrow \frac{4}{\sigma_\delta^8} \ln \lp 1 + \sigma_\delta^2\rp.
\end{equation}
 Both series given in (\ref{fmoments}) are quickly convergent and well suited for numerical evaluation. Figure \ref{fig1} shows the accessible fractions $\epsilon_a$ of information through extraction of the full series moments. Figure \ref{fig2} shows the repartition of this accessible fraction among the first 10 moments. Most relevant from a cosmological point of view in figure \ref{fig1} is the solid line, dealing with the case of the  parameters of interest entering the variance only. These figures shows clearly that the moments, as probes of the lognormal matter field, are penalized by two different processes. First, as soon as the field shows non-linear features, following equations (\ref{asymptotics}) and (\ref{asymptotics2}), almost the entirety of the information content cannot be accessed anymore by extracting its successive moments. Within a range of one magnitude in the variance, the moments goes from very efficient probes to highly inefficient. Second, as shown in figure \ref{fig2}, as  the variance of the field approaches unity, this accessible fraction gets quickly transferred from the variance alone to higher order moments. This repartition of the information within the moments is built out of two different regimes. First, for large variance, or  large $n$, we see easily from the above expressions (\ref{anbn}) and (\ref{Ansigma}) that in both cases the information coefficients decays exponentially,
\begin{equation} \label{as1}
s_n^2 \propto \lp 1 + \sigma_\delta^2 \rp^{-n},\quad n\ln \lp 1 + \sigma^2_\delta \rp  \gg   1.
\end{equation}
On the other hand, if the variance or $n$ is small enough, we can set $1 - q^n \approx -n\ln q $, and we obtain, very roughly,
\beq\label{as2}
s_n^2 \propto \lb n \ln \lp 1 + \sigma_\delta^2 \rp\rb^{n},\quad n\ln \lp 1 + \sigma^2_\delta \rp   \ll 1,
\enq
explaining the trend with variance seen in figure \ref{fig2}, that puts more importance to higher order moments as the variance grows.
Note that the latter regime can occur only for small enough values of the variance. Deeper in the non linear regime, the trend is therefore reversed, obeying (\ref{as1}) for all values of $n$, with a steeper decay for higher variance.

 \subsection{A $q$-analog of the logarithm}
 \begin{figure}
  \includegraphics[width = 0.5\textwidth]{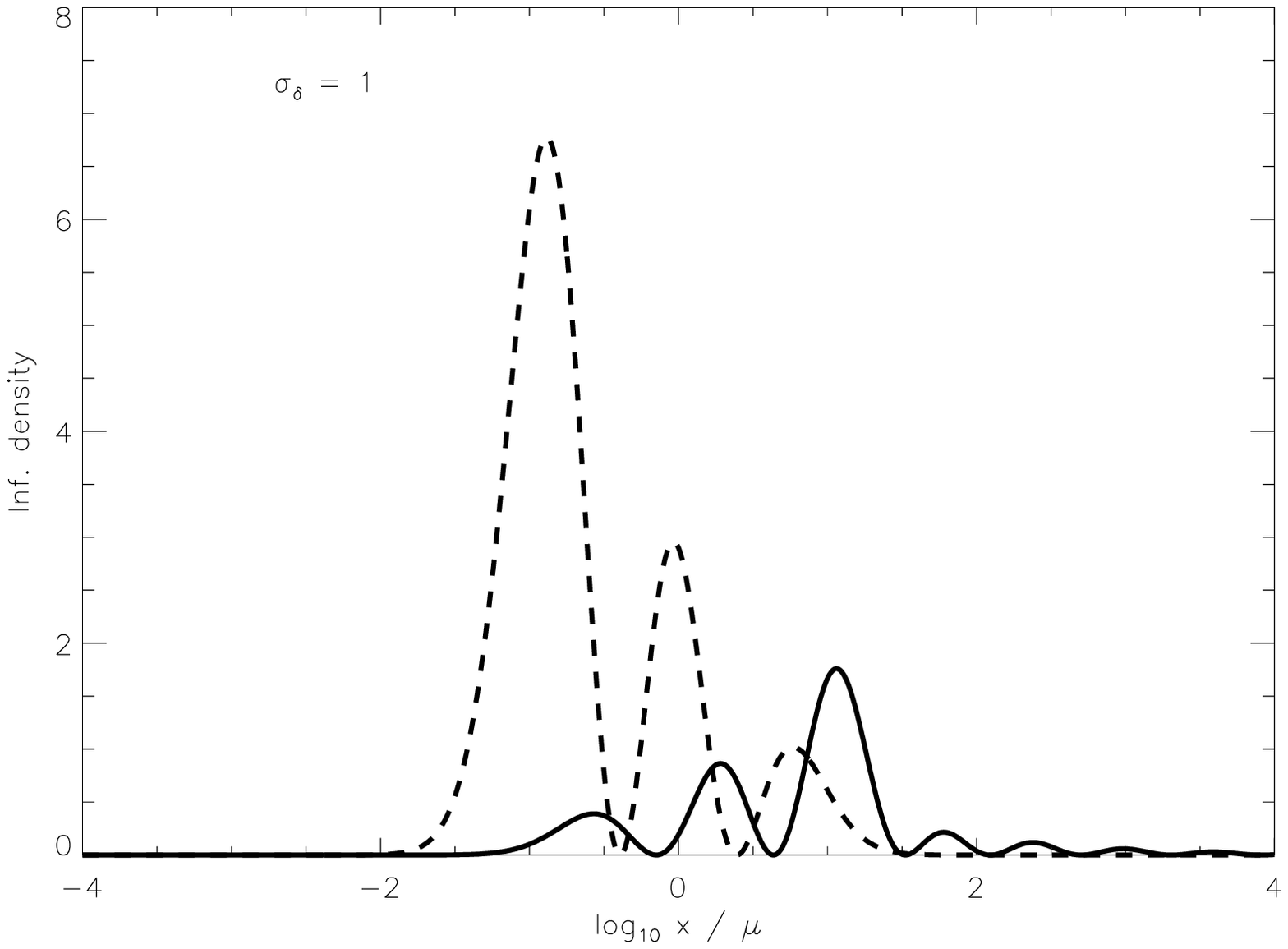}
 \caption{\label{fig3}The information density of the lognormal distribution, dashed, and, solid, its approximation through the associated orthonormal polynomials, in the $a = \sigma$ case, for fluctuations of unit variance. While most of the information of the lognormal field in this regime is actually contained in the underdense regions, the moments are essentially unable to catch it.}
 \end{figure}
These results show clearly that large parts of the information become invisible to the moments. However, it does not tell us what is responsible for this phenomenon. It is therefore of interest to look in a little bit more into the details of this missing pieces of information. As we have seen, these are due to the inability of the polynomials to reconstruct precisely the score function. 
In the case $a = \mu$, the score function of the lognormal distribution is easily shown to take the form of a logarithm in base $q$,
\beq
\begin{split}
s(x)
& = -\frac 12 -  \ln_q \lp \frac  {x} {\mu}\rp .
\end{split}
\enq
Therefore the series
\beq
s^\mu(x) := \sum_{n = 0}^\infty s_n^\mu P_n(x)
\enq
will represent some function, very close to a logarithm for $q \rightarrow 1$ over the range of $p(x,\alpha)$. It will however fail to reproduce some of its features at lower $q$-values. This is hardly surprising, since it is well known that the logarithm function does not have a Taylor expansion over the full positive axis. For this reason, the approximation $s^\mu(x)$ of $s(x)$ through polynomials can indeed only fail when the fluctuation variance becomes large enough. In the appendix, we show that $s^\mu(x)$ takes the form
\beq \label{smu}
s^\mu(x) = - \sum_{k = 1}^\infty \frac{q^k}{1-q^k}\lb 1 +(-1)^k\frac{q^{k(k-1)}}{\pochh  q q k} \lp \frac{x}{\mu}\rp^k\rb.
\enq
 It is interesting to note that this series expansion is almost identical to the one of the $q$-analog of the logarithm $S_q(x)$  defined by E. Koelink and W. Van Assche, with the only difference being the replacement of $q^{k(k-1)/2} $ by $q^{k(k-1)}$ (See \citep{Koelink09}, and also \citep{Gautschi06}). Due to this replacement, $s^\mu$ does not possess several properties $S_q$ has and makes it a real $q$-analog of the logarithm, such as $S_q(q^{-n}) =  n$, for positive integers. The qualitative behavior of $s^\mu$ stays however close to $S_q$. Notably, its behavior in underdense regions, $x/ \mu \ll 1$,  where as seen from (\ref{smu}) $s^\mu$ tends to a finite value, is very different from a logarithm.
 \newline\newline
This calculation can be performed as well in the case $a = \sigma$, with similar conclusions. Since it is rather tedious and not very enlightening, we do not reproduce it in these pages. We show in figure \ref{fig3} the information density, equation (\ref{infdensity}), of the lognormal distribution (dashed line), 
 and its approximation by the orthogonal polynomials (solid line),
 \beq
 p(x,\alpha)\lp \sum_{n = 0}^\infty s_n^\sigma P_n(x)\rp^2,
 \enq
 when the fluctuation variance $\sigma^2_\delta$ is equal to unity. It is clear from this figure that in this regime, while most of information is located within the underdense regions of the lognormal field, the moments are however unable to catch it. \newline
 To check the correctness of our numerical and analytical calculations, we compared the total information content as evaluated from integrating the information densities on figure \ref{fig3} to the one given by the equation (\ref{Fsigex}), respectively (\ref{fmoments}), with essentially perfect agreement.
\subsection{Comparison to standard perturbation theory}
\begin{figure}
  \includegraphics[width = 0.5\textwidth]{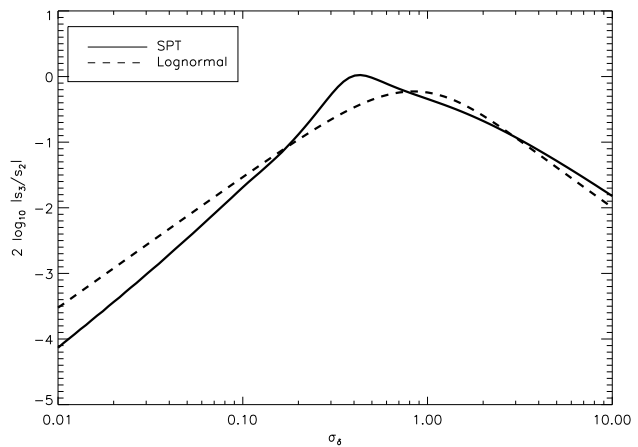}
 \caption{\label{fig4}The ratio of the independent information content of the third moment to that of the second moment, for the lognormal field (dashed) and standard perturbation theory (solid), as function of the square root of the variance of the fluctuations. }
 \end{figure}

For any distribution, the knowledge of its first $2n$ moments allow directly, for instance from equation (\ref{infn}), the evaluation of the independent information content of the first $n$ moments. This even if the exact shape of the distribution is not known, or too complicated. In particular, we can use the explicit expressions for the first six moments of the density fluctuation field within the framework of standard perturbation theory (SPT) provided by F. Bernardeau in \citep{1994A&A...291..697B}, in order to compare $s_2$ and $s_3$ as given from SPT to their lognormal analogs.\newline\newline
We note that a comparison to \citep{1994A&A...291..697B} can only be very incomplete and, to some extent, it can only fail. It is indeed part of the  approach in \citep{1994A&A...291..697B}, when producing functional forms for the distribution of the fluctuation field, to invert the relation between a moment generating function and its probability density function. For such an inversion to be possible it is of course necessary that the probability density is uniquely determined by its moments. As said, this is not the case for the lognormal distribution. Therefore, that approach can never lead to an exact lognormal distribution, or to any distribution for which the moment hierarchy forms an incomplete set of probes. However, such a comparison can still lead to conclusions relevant for many practical purposes, such as those dealing with the first few moments.
\newline\newline
The variance of the field is explicitly given as an integral over the matter power spectrum,
\beq \label{varbdeau}
\sigma^2_\delta  = \frac{1}{2\pi^2}\int_0^\infty dk \:k^2P(k,\alpha) \left|W(kR)\right|^2,
\enq
where $W(kR)$ is the Fourier transform of the real space top hat filter of size $R$, and any cosmological parameter $\alpha$ entering the power spectrum $P(k)$.
In the notation of \citep{1994A&A...291..697B}, the moments of the fluctuation field $m_n = \av{\delta^n}$ are given by the deconnected, or Gaussian, components, while the connected components $\av{\delta^n}_c ,\quad n \ge 3$ are given in terms of parameters $S_n$,
\beq \label{mbdeau} \begin{split}
\av{\delta^n}_c = \sigma^{2(n-1)}S_n.
\end{split}
\enq
The parameters $S_n$ contain a leading, scale independent coefficient, and deviation from this scale independence are given in terms of the logarithmic derivative of the variance,
\beq
\gamma_i = \frac{d^i \ln \sigma^2_\delta}{d\:\ln R^i},\quad i = 1,\cdots
\enq
Neglecting the very weak dependence of $S_n$ on cosmology, from (\ref{mbdeau}) we can write
\beq
\begin{split}
 \frac{\partial m_n}{\partial \alpha} = \frac{\partial \sigma_\delta^2}{\partial \alpha}\cdot \begin{cases} 0,&\quad n = 1 \\  1,&\quad n = 2
 \\ 2 m_3 \:/ \: \sigma_\delta^2,&\quad n = 3  \end{cases}
\end{split}
\enq
With the coefficients $S_n$ up to $n = 6$ given in  \citep[page 703]{1994A&A...291..697B}, and the above relations, we performed a straightforward evaluation of the information coefficients $s_2^2$ and $s_3^2$, using equation (\ref{infn}). The variance was obtained from (\ref{varbdeau}) within a flat $\Lambda$CDM universe ($\Omega_\Lambda = 0.7,\Omega_m = 0.3,\Omega_b = 0.045, h = 0.7$),  with power spectrum parameters ($\sigma_8 = 0.8, n = 1$) and we used the transfer function from Eisenstein and Hu \citep{1998ApJ...496..605E}. The needed derivatives $\gamma_i, \: i = 1,\cdots, 4$ were obtained numerically through finite differences.
\newline\newline
In figure \ref{fig4}, we show the ratio
\beq
\lp \frac{s^\sigma_3}{s^\sigma_2} \rp^2
\enq
i.e. the relative importance of the third moment with respect to the second, as function of the variance, both for the lognormal distribution and the SPT predictions. This ratio is identically zero for a Gaussian distribution.
The models stands in good agreement over many orders of magnitude. It is striking that both models consistently predict that a the entrance of the non-linear regime,  this ratio takes a maximal value close to unity. Surely, the SPT curve for larger values of the variance is hard to interpret, since out of its domain of validity. 
\section{Several variables}
\newcommand{\vecn}{\mathbf n}
\newcommand{\vecm}{\mathbf m}
\newcommand{\veci}{\mathbf i}
\newcommand{\vecj}{\mathbf j}
So far, we have considered only one variable, and wish now to extend our analysis to the more interesting multidimensional case. It is possible to derive formally a general expression for the independent information content of the $n$-point correlations of any distribution. This proceeds in strict analogy with the one-dimensional case, where an expansion of the score function in polynomials of several variables is made. It is presented in some more details in the appendix. For the lognormal field, a given model parameter can only enter via the means of the logarithm of the field at each point or through the elements of its two-point correlation matrix.  We could however not transform the corresponding expressions in a useful, easily evaluated form in a general situation, like in the one dimensional case.  We focus therefore on two more restricted but tractable situations. First, in analogy with the $a = \mu$ case, we consider a parameter that enters the mean of the field $\rho $ and no elements of the correlation matrix. In this case, the complete amount of information can be extracted via the mean of $\ln \rho$, and compare that amount to the one obtained by extracting the mean of $\rho$ only. In a second step, we consider the extraction of the correlation amplitude $\xi(r)$ between independent pairs of cells separated by that distance $r$. In this case, the full amount of information on any parameter impacting $\xi(r)$ is in the correlation of $\ln \rho$, and compare that to the more standard approach of the extraction of the correlation of $\rho$.
\subsection{Basic properties of the multivariate lognormal distribution}
The random vector $\rho = \lp \rho_1,\cdots,\rho_d \rp$ has a multivariate lognormal distribution if $\ln\rho := \lp \ln \rho_1,\cdots, \ln \rho_d \rp$ is normally distributed, for some
 mean vector $\bar \ln\rho$ and covariance matrix $\xi_{\ln \rho}$. 
 Since we have in mind the vector $\rho$ to be a sample of an homogeneous lognormal field, we will use the notation
 \beq
 \rho_i = \rho(x_i),\quad i = 1,\cdots,d,
 \enq
 the points $x_i$ lying in some $n-$dimensional space. The means $\mlnrho_i$ are the same at each point, and the covariance matrix is  a discrete version of the corresponding correlation function,
 \beq
 \lb \xi_{\ln \rho} \rb_{ij} = \xi_{\ln\rho}(x_i - x_j),\quad i,j = 1,\cdots,d
 \enq
 which depends only on the separation vector.
 For some vector of natural numbers  (multiindex) $\vecj = (j_1,\cdots,j_d)$, the correlations
 \beq
 m_\vecj = \av{\rho^\vecj }=  \av{\rho(x_1)^{j_1}\cdots\rho(x_d)^{j_d}}
 \enq
 are again given by simple Gaussian integrals and read explicitly
 \beq \label{moments2}
 m_\vecj = \exp\lp \mlnrho \cdot \vecj + \frac 12 \:\vecj^T \xi_{\ln \rho} \cdot \vecj \rp.
 \enq
 By the independent information content $s_n^2$ of the n-point correlations we mean the independent bits of information within all the correlations of the same order $n$, that is within all $m_\vecj$ such that
 \beq
 |\vecj| := \sum_{i = 1}^d j_i = n.
 \enq
The convenient property (\ref{separability}) between the moments becomes
 \beq \label{separability2}
 m_{\veci + \vecj} = m_\veci \:m_\vecj \exp \lp \veci ^T \xi_{\ln \rho} \cdot \vecj \rp :=  m_\veci \:m_\vecj \:Q_{\veci\vecj}.
 \enq
 From these relations, one infers that the correlations of the fluctuations $\xi$ of $\rho$, defined as
 \beq \label{xidef}
  \xi(x_i-x_j) = \frac{1}{\mrho^2}\av{ \lp \rho(x_i) - \mrho \rp\lp \rho(x_j) - \mrho \rp}, 
 \enq
  are related to those of $\ln\rho$ through
 \beq \label{corrrel}
 \begin{split}
 \xi_{\ln \rho}\lp r \rp & = \ln\lb 1 + \xi \lp r \rp \rb.
 \end{split}
 \enq
 On the other hand. the means obey
 \beq
 \begin{split}
  \mlnrho &= \ln \mrho-\frac 12  \xi_{\ln \rho}(0) \\ 
  &=  \ln \mrho-\frac 12  \ln \lp 1 + \sigma^2_\delta \rp.
  \end{split}
 \enq
 The Fisher information content of $\rho$ is again given by the standard expression for the Gaussian field $\ln\rho$. It splits into the part coming from the observation of $\mlnrho$ and the one coming from the correlations $\xi_{\ln\rho}$,
 \beq\label{infmulti}
 F_\alpha = \frac 12 \Tr \lb \xi_{\ln \rho}^{-1} \frac{\partial \xi_{\ln \rho}}{\partial \alpha}\:\xi_{\ln \rho}^{-1} \frac{\partial \xi_{\ln \rho}}{\partial \alpha} \rb + \frac{\partial \mlnrho}{\partial \alpha}\xi_{\ln\rho}^{-1} \cdot \frac{\partial \mlnrho}{\partial \alpha}.
 \enq
 We denote by $Q_N$ the square matrix defined as
\beq
\lb Q_N  \rb_{\veci\vecj} = Q_{\veci\vecj},\quad |\veci|,|\vecj| \le N,
\enq
which we will make future use of. We note that by virtue of equations (\ref{separability2}) and (\ref{corrrel}), its matrix elements read
\beq \label{qm}
Q_{\veci\vecj} = \prod_{k,l = 1}^d\lp 1 + \xi \lp x_k - x_l\rp \rp^{i_kj_l}.
\enq
A general expression for the independent information content $s_n^2$ of the correlations of order n, equation (\ref{smulti}), and its link to the completeness of the orthogonal polynomials is presented for completeness in the appendix. This machinery is however not compulsory for the following considerations, which are restricted to the two lowest order correlations. We will only use the analog of equation (\ref{infn}), which gives the total amount of information contained in the correlations up to order $N$,
\beq\label{infm}
\sum_{n = 1}^N s_n^2 = \sum_{|\veci|,|\vecj| = 1}^N \frac{\partial m_\veci}{\partial \alpha}\lb \Sigma^{-1} \rb_{\veci\vecj}\frac{\partial m_\vecj}{\partial \alpha},
\enq
where $\Sigma$ is the covariance matrix
\beq
\Sigma_{\veci\vecj} = m_{\veci + \vecj} - m_\veci m_\vecj,\quad|\veci|,|\vecj| \le N.
\enq
For the lognormal distribution, the property (\ref{separability2}) allow us to rewrite this last expression in the equivalent form
\beq \label{infsum}
\sum_{n = 1}^N s_n^2 = \sum_{|\veci|,|\vecj| = 0}^N \frac{\partial \ln m_\veci}{\partial \alpha}\lb  Q_N^{-1} \rb_{\veci\vecj}\frac{\partial \ln m_\vecj}{\partial \alpha}.
\enq
\subsection{Extraction of the mean}
In this case, we set
\beq \label{cond}
\frac{\partial \xi(r)}{\partial \alpha } = 0
\enq
for any argument. Again, the actual value of the derivative of the mean with respect to $\alpha$ will play no role.
This condition (\ref{cond}) implies that $\partial_\alpha \xi_{\ln \rho} $ vanishes as well. We can read out from equation (\ref{infmulti}) that the total amount of information is given by
\beq
\begin{split}
F_\alpha &= \lp \frac{\partial \ln \mrho}{\partial \alpha}\rp^2 \sum_{i,j = 1}^d \lb \xi_{\ln\rho}^{-1}\rb_{ij} \\
\end{split}
\enq
We stress that since $\ln\rho$ is Gaussian, the information is accessible in its entirety by extraction of the mean of $\ln \rho$. On the other hand, the amount extracted by looking at the mean of $\rho$ itself is equation (\ref{infm}) with $N = 1$.  Using the definition of $\xi$ in equation (\ref{xidef}), it becomes
\beq
s_1^2 =  \lp \frac{\partial \ln \mrho}{\partial \alpha}\rp^2 \sum_{i,j = 1}^d \lb \xi^{-1} \rb_{ij}.
\enq
In the limit of a continuous sample  $d \rightarrow \infty$, the sum
\beq
\sum_{i,j = 1}^d \lb \xi^{-1} \rb_{ij}
\enq
becomes a double integral over space, which can be performed by Fourier transformation, and is the inverse of the power spectrum of the field at zero argument,
\beq \begin{split}
\sum_{i,j = 1}^d \lb \xi^{-1} \rb_{ij} & \stackrel{N\rightarrow \infty}{\rightarrow} \frac{ V}{ P_\rho( k = 0)} \\
P_\rho(0) &= \int_V d^nr \:\xi(r) 
\end{split}
\enq
and similarly for $\xi_{\ln\rho}$. We conclude that the loss of information by looking at the mean of the field only is given straightforwardly by the ratio of the power of the fields $\ln \rho$ and $\rho$ at zero argument.
\beq
\epsilon := \frac{s_1^2}{F_\alpha} =\frac{P_{\ln \rho}(0)}{P_\rho(0)}.
\enq
From the explicit representation of $P_{\ln\rho}(0)$, 
\beq
P_{\ln \rho}(0) = \int_V d^nr \ln \lp 1 + \xi(r)\rp
\enq
and the fact that $\ln \lp 1 + x \rp$ is strictly smaller than $x$ whenever $x \ne 0$, it follows that the loss of information always occurs, but is substantial only if the correlation function takes substantial values. However the information loss $\epsilon$ is roughly insensitive to the presence of some correlation scale. We see that the presence of correlations does not alter the main conclusions drawn in the first part of his work. \newline
\subsection{Extraction of correlations}
\begin{figure}
  \includegraphics[width = 0.5\textwidth]{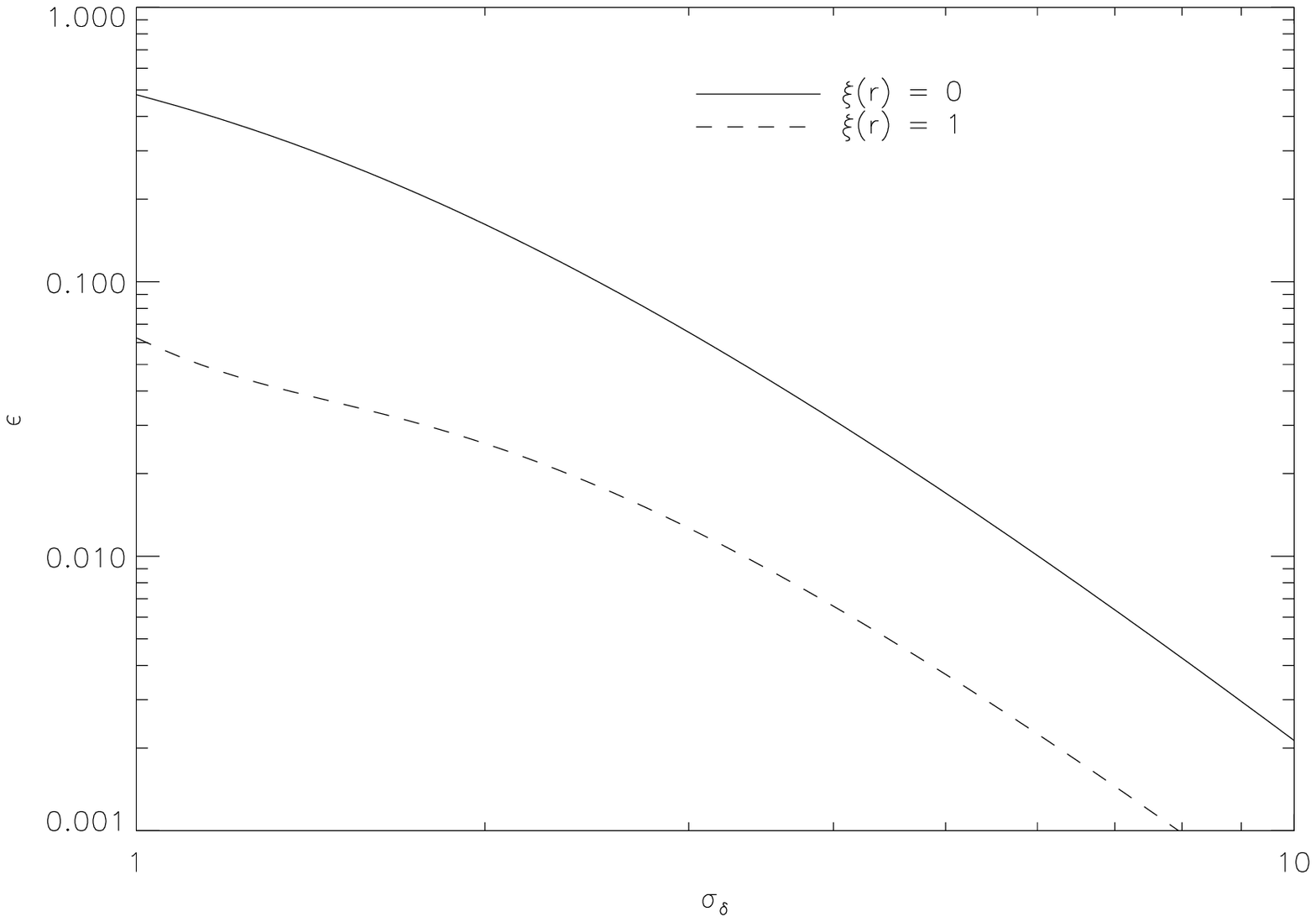}
 \caption{\label{figxi} The loss of information $\epsilon$ in extracting correlations of $\rho$ instead of $\ln \rho$ for a lognormal field, equation (\ref{infoxiloss}), in the limit of vanishing correlations (solid line), and at the non linearity scale $\xi(r) = 1$ (dashed) as function of the square root of the variance of the fluctuations.}
 \end{figure}

We suppose now the parameter $\alpha$ enters the correlation function $\xi$ for some argument $r$.
Since the field is lognormal, measurement of $\xi_{\ln\rho}(r)$ captures all the information on $\alpha$. We want to compare this amount to the one extracted by measuring $\xi(r)$ itself. We suppose further that $\xi(r)$ is extracted from a number of independent pairs of points separated by that distance $r$. The independency of the pairs allow us to simplify drastically the problem, since by additivity of the information the information loss will be  independent of the number of pairs. Our problem becomes thus two-dimensional. Our assumptions on the impact of the parameter $\alpha$ are, more explicitly, 
\beq \label{ass2}
\frac{\partial \mrho}{\partial \alpha }= 0,\quad \frac{\partial \sigma^2_\delta}{\partial \alpha} = 0,\quad \frac{\partial \xi(r)}{\partial \alpha} \ne 0.
\enq
We point out that this is very different from the $a = \sigma$ case that we treated earlier, since here the variance $\sigma^2_\delta$ only acts as a noise source and not as a source of information.
The correlation matrix of a pair of points of the homogeneous Gaussian field $\ln \rho$ separated by $r$ reads, according to (\ref{corrrel}),
\beq
\xi_{\ln \rho} = \begin{pmatrix} \ln \lp 1 + \sigma^2_\delta \rp & \ln  \lp 1 +\xi(r) \rp  \\ \ln \lp 1 +\xi(r) \rp & \ln  \lp 1 + \sigma^2_\delta \rp & \end{pmatrix}.
\enq
The positivity of the matrix constrains, at a fixed variance $\sigma^2_\delta$, the values of $\xi(r)$ to the following range
\beq
1 + \xi(r) = \lp 1 + \sigma^2_\delta\rp^\eta,\quad \eta \in (-1,1).
\enq
Clearly, vanishing correlations corresponds to $\eta = 0$, positive correlations to $\eta > 0$, and negative correlations to $\eta < 0$. 
By assumption, the parameter $\alpha$ does enter $\xi(r)$ only. Therefore,
\beq \label{above}
\frac{\partial \xi_{\ln \rho}}{\partial \alpha} = \frac{\partial_\alpha \xi (r)}{1 + \xi(r)} \begin{pmatrix} 0 & 1 \\ 1 & 0 \end{pmatrix}.
\enq
Putting (\ref{above}) in equation (\ref{infmulti}), we obtain that the total amount of information on $\alpha$ is
\beq\label{f2ptln}
F_\alpha = \lp \frac{\partial_\alpha \xi (r)}{1 + \xi(r)}\rp^2\frac{1 + \eta^2}{\lp 1 - \eta^2 \rp^2}\frac{1}{\ln^2\lp 1 + \sigma^2_\delta \rp}.
\enq
To obtain the information obtained by extracting $\xi$, we first note that under our assumption (\ref{ass2}) the mean carries no information,
\beq
s_1 = 0.
\enq
For this reason, $s_2^2$ is given by equation (\ref{infsum}), with $N = 2$. The only non-zero element of the vector of derivatives $\partial_\alpha \ln m_\veci,\:|\veci| \le 2$, is in our configuration (\ref{ass2}) for the multiindex  $\veci = (1,1)$. From  equations (\ref{moments2}) and (\ref{corrrel}), we obtain
\beq
\frac{\partial \ln m_{(1,1)}}{\partial \alpha } = \frac{\partial \xi(r)}{\partial \alpha}\frac 1 {1 + \xi(r)}.
\enq
It follows immediately that $s_2^2$ is given by
\beq\label{f2pt}
s_2^2 = \lp \frac{\partial_\alpha \xi (r)}{1 + \xi(r)}\rp^2\lb Q_2^{-1} \rb_{(1,1)(1,1)}.
\enq
 Of special interest is the limit of low correlations, where the exact result 
\beq \label{xi0}
\frac{s_2^2}{F_\alpha} = \frac1  {\sigma^4_\delta} \ln^2\lp 1 + \sigma^2_\delta\rp, \quad \textrm{for}\quad \xi(r) = 0.
\enq
can be obtained making profit of the simple structure of the $Q_2$ matrix.
We note that $Q_2$ is in our two-point configuration a 6 dimensional matrix, where, as seen from its representation (\ref{qm}), all of its elements are products and powers of $1 + \sigma_\delta^2$ and $1 + \xi(r)$.
The needed inverse matrix element can be written as ratio of determinants
\beq \label{poly}
\lb Q_2^{-1} \rb_{(1,1)(1,1)} = \frac{\det \hat Q_2}{\det Q_2}, 
\enq
where $\hat Q_2$ is the 5-dimensional matrix originating from $Q_2$ where the row and column corresponding to the multiindex $(1,1)$ have been taken out. Both of these determinants are therefore clearly polynomials in $1 + \sigma_\delta^2$ and $1 + \xi(r)$. The asymptotic behavior of the accessible information for large variance can be thus obtained by noticing that $\eta \rightarrow 0$ for any value of $\xi$. Looking then at the  leading coefficients of the two polynomials entering (\ref{poly}), we obtain
\beq 
\begin{split}
\det Q_2 & \rightarrow \lp \sigma^2_\delta \rp^{12}\lp 1 + \xi(r) \rp^2\\
\det \hat Q_2 & \rightarrow \lp \sigma^2_\delta \rp^{10}.
\end{split}
\enq
Therefore, the information loss
\beq \label{infoxiloss}
\epsilon := \frac{s_2^2}{F_\alpha}
\enq
tends to, for asymptotic values of the variance,
\beq\begin{split}
\epsilon &\rightarrow \frac 1 {\sigma^4_\delta}  \frac{\ln^2\lp 1 + \sigma^2_\delta  \rp} {\lp1 + \xi(r) \rp^2} \\
&= \frac{\epsilon (\xi = 0)} {\lp1 + \xi(r) \rp^2}.
\end{split}
\enq
The second line follows from the first using the exact result for vanishing correlations given in equation (\ref{xi0}).
The efficiency of $\xi$ in extracting the information on any parameter therefore always goes to zero, following approximately the inverse squared variance. The presence of substantial positive correlations, generic for a field generated by gravitational instability on a wide range of scales, only makes the information loss worse.
This is illustrated in figure \ref{figxi}, where the dotted line shows the loss of information at the non linearity scale $\xi(r) = 1$, evaluated numerically from equations (\ref{f2ptln}) and (\ref{f2pt}), together with the exact result (\ref{xi0})  for vanishing correlations (solid line).

\section{Summary and conclusion}
We have investigated in details the structure of the information within the moments of the univariate lognormal distribution, as a model for the matter density field. We have provided exact expressions, equations (\ref{anbn}) and (\ref{Ansigma}) for the independent information content of each moment. Using these expressions, we have shown that the moments become dramatically incomplete probes in the non linear regime.  In the cosmologically relevant case of the parameter entering the power spectrum of the fluctuations, the fraction of the information that is accessible from the moments is close to one fourth at $\sigma_\delta = 1$, and decays following the inverse $4$th power of the variance. We showed it is mainly due to the inability of the moments to probe the information located in the underdense regions of the distribution. Besides, a comparison to standard perturbation theory showed for the lower order moments showed that both approaches are consistent and predict that the third order moment becomes as important as the variance itself when entering the non linear regime.\newline\newline In a second step, we extended our results to the multivariate case, showing that the mean of the field and two-point correlations become in a very similar manner very inefficient probes, for any parameter of interest. With the help of two simplified situations we have shown that the presence of correlations only makes the information loss even worse. More specifically, we have shown that  the extraction of the two-point correlation function at any argument provides access to a fraction of the information which is generically well below unity at the entrance of the non linear regime, and decays like the inverse squared variance. 
\newline\newline
These results, making clear that the information content of the lognormal field not only gets transferred to higher order point functions, but also becomes largely inaccessible to the correlation function hierarchy in the non linear regime, confirm to full extent qualitative suspicions raised in \citep[section 4]{1991MNRAS.248....1C}, to which we refer for a more complete discussion on physical arguments that may source such a behavior. \newline\newline
We can understand for any random field if the hierarchy members are promising probes or not from the following considerations. Let $p[\phi,\alpha]$ be the probability density function for a realization $\phi$ of the field. As we have seen, the information content of the first $N$-correlation functions is based on the approximation of the score function $\partial_\alpha \ln p[\phi] $ through polynomials up to order $N$, over the range of $p[\phi]$. Let us assume for instance that for any value of the model parameter  $\ln p[\phi]$ can be expanded in a low order Taylor series in the field,
\beq \label{expfield}\begin{split}
\ln p[\phi,\alpha] &= \sum_{n = 0}^N \int dx_1\cdots dx_n \\
&\lambda_n \lp x_1,\cdots,x_n,\alpha \rp \phi(x_1)\cdots\phi(x_n). 
\end{split}
\enq
This class of distributions, including Gaussian fields for which $N = 2$, can arise notably as maximal entropy distributions for fixed values of the first $N$-correlation functions. The coefficient $\lambda_n$ is called in this framework the potential associated to the $n$-th correlation function (See \citep[e.g.]{jaynes83,caticha08}). Since $\partial_\alpha \ln p$ is itself of polynomial of order $N$, all the information is contained is the first $N$-correlation functions. Of course, the relative importance of each one of these will be modulated by the sensitivity of the potentials to the parameter $\alpha$ and their covariances. This situation is certainly the one where the correlation function hierarchy are the probes of choice, since only a finite number of these grasp the entire information content. \newline\newline
Two different processes may at this point render the hierarchy inefficient, or incomplete. First of all, if a large number of terms are needed in the expansion (\ref{expfield}) to reproduce accurately the score function $\partial_\alpha \ln p$. In this case, one would need to go deep down the hierarchy in order to catch the information. This is certainly not desirable. The last case occurs when $\partial_\alpha \ln p$ has no Taylor expansion at all over the relevant range. It is then simply not possible to represent accurately the score function. Parts of the information (given in the field analog of equation (\ref{MSE})) becomes invisible to the correlation function hierarchy. The lack of a Taylor expansion for the logarithm function is the reason for the failure of the moments and correlation functions to catch the information of a lognormal field in the non linear regime, when the range of the probability density function becomes very large. \newline\newline
We emphasize, as in \citep{1991MNRAS.248....1C}, that these peculiar dynamics of the information are not due to a pathological character of the lognormal distribution. It should be expected for any distribution decaying slowly at infinity. We can add to their discussion that this is so because  $\ln p$ cannot be well reproduced by polynomials under this condition.
\newline\newline
Given the very high amplitude of these effects within the lognormal assumption, we believe that in order to get the best out of future galaxy survey data, it is crucial to understand better these issues. The present work presents first steps to this aim. It is also very consistent and brings strong support to the recent studies started in \citep{2009ApJ...698L..90N,2011ApJ...731..116N,2011arXiv1103.2166W} making in the non linear regime the logarithm of the field rather than the field itself the central quantity of interest. It remains however to be seen to what extent the approach presented in this work is able to provide quantitative predicitions for the statistical power of higher point functions, or for power spectrum extraction. We leave these aspects for future work.

\acknowledgments
We would like to thank Adam Amara, Simon Lilly, Alexander Szalay and Mark Neyrinck for useful discussions, and acknowledge the support of the Swiss National Science Foundation.



\appendix

\section{Derivation of relation (\ref{lemma}) \label{proof}}
To prove (\ref{lemma}),  we note that both sides of the equation are polynomials of degree $n$ in $t$, and that the zeroes of the right hand side are given by
\beq
t = q^{-i}, \quad i = 0,\cdots, n-1.
\enq 
We first show that the left hand side evaluated at these points does vanish as well, so that the two polynomials must be proportional. We then find the constant of proportionality by requiring $P_n$ to have the correct normalization. \newline
The first step is performed by noting that
\beq
\av{P_n(q^{-i}x)} = \frac 1 {m_i}\av{P_n(x)x^i }, \quad i = 0,1,\cdots
\enq
an identity which is proven by expanding $P_n $ in both sides of the equation  in terms of the transition matrix $C$, and using the relation (\ref{separability}) between the moments.
Since $P_n$ is by construction orthogonal to any polynomial of strictly lower degree, we have indeed
\beq
\av{P_n(q^{-i}x)} = 0,\quad i = 0,\cdots,n-1. 
\enq
This implies
\beq \label{sumrule}
\sum_{k = 0}^n C_{nk} m_k\:t^k  =\alpha_n \pochh t q n
\enq
for some constant of proportionality $\alpha_n$. To find it, 
we note that by expanding the normalization condition of $P_n$,
\beq \begin{split}
1 &= \av{P_n^2(x)}, 
\end{split}
\enq
using again property (\ref{separability}), it must hold that
\beq
1 = \sum_{i,j = 0}^nC_{ni}m_j\: C_{nj}m_j\: q^{-ij}.
\enq
The sums can be performed using equation (\ref{sumrule}), leading to the following equation for $\alpha_n$,
\beq
1 = (-1)^n\:\alpha_n^2\: q^{n(n-1)/2}\pochh {q^{-n}}qn.
\enq
This expression simplifies to
\beq
\alpha_n^2 = \frac{q^n}{\pochh q q n}
\enq
and the sign of $\alpha_n$ must be $-1^n$ in order to have a positive matrix element $C_{nn}$. This concludes the proof of (\ref{lemma}).
\section{Derivation of  the representation (\ref{smu})}
In order to get the explicit series representation of (\ref{smu}), we first obtain from relation (\ref{lemma}) the exact expression of the transition matrix $C$. The expansion of the $q$-Pochammer symbol on the right hand side of (\ref{lemma}) in powers of $t$ is the Cauchy binomial theorem,
\beq
\pochh t q n = \sum_{k = 0}^n \binq n k q q^{k(k-1)/2}(-t)^k,
\enq   
where
\beq
\binq n k q = \frac{\pochh q q n}{\pochh q q k \pochh q q {n-k}}
\enq is the Gaussian binomial coefficient. Matching powers of $t$ in (\ref{lemma}) we obtain the explicit form
\beq \label{}
C_{nk}=(-1)^{n-k} \frac{q^{n/2}}{\sqrt{\pochh q q n}} \binq n k q q^{k^2}\mu^{-k}.
\enq
Therefore, interchanging the $n$ and $k$ sums in (\ref{smu}) , it holds
\beq
s^\mu(x) =-\sum_{n = 1}^\infty \frac{q^n}{1- q^n} + \sum_{k = 1}^\infty q^{k^2} \lp -\frac x  {\mu}\rp^k \sum_{n = k}^\infty\frac{q^n}{1-q^n}\binq n k q.
\enq
With the help of some algebra the following identity is not difficult to show
\beq
\sum_{n = k}^\infty\frac{q^n}{1-q^n}\binq n k q = \frac 1 {\pochh q q k}\frac{q^k}{1 - q^k},\quad k \ge 1.
\enq
Consequently, the series expansion of $s_\mu(x)$ is given by
\beq
s^\mu(x) = - \sum_{k = 1}^\infty \frac{q^k}{1-q^k}\lb 1 +(-1)^k\frac{q^{k(k-1)}}{\pochh  q q k} \lp \frac{x}{\mu}\rp^k\rb
\enq
\section{Several variables}
We first need a little bit of notation. For a variable $X$ taking values $x = \lp x_1,\cdots x_d\rp$, we use the multindex notation
\beq
\begin{split}
x^\vecn &= x_1^{n_1}\cdots x_d^{n_d} \\
\vecn &= \lp n_1,\cdots n_d\rp,\quad n_i = 0,1,\cdots. \\
|\vecn| &:= \sum_{i = 1}^dn_i,
\end{split}
\enq
where $|\vecn|$ is the order of the multiindex $\vecn$.
A moment of order $N$ is given by
\beq
m_\vecn = \av{x^\vecn}, \quad |\vecn| = N,
\enq
and the covariance between the moments is
\beq
m_{\vecn + \vecm} - m_\vecn m_\vecm =: \Sigma_{\vecn\vecm}.
\enq
In this notation, the decomposition of the information in independent bits of order $N$ proceeds by strict analogy with the one dimensional case. We refer to \citep{Dunkl01} for the general theory of orthogonal polynomials in several variables. A main difference being that at a fixed order $N$ there are not one but $\begin{pmatrix} N+ d -1\\  d \end{pmatrix}$ independent orthogonal polynomials, which are not uniquely defined. The orthogonality of the polynomials of same order is not essential for our purposes, but requiring the following condition is enough,
\beq \label{conditions}
\begin{split}
\av{P_\vecn(x) P_\vecm(x)} &= 0 ,\quad |\vecn| \ne |\vecm| \\
\av{P_\vecn(x) P_\vecm(x)} &= \lb H_n\rb_{\vecn\vecm} ,\quad |\vecn| = |\vecm| = n\\
\end{split}
\enq
for some matrices $H_n$.
The component of some function $f$ parallel to the polynomial $P_\vecn$ is
\beq
s_\vecn := \av{f(x)P_\vecn(x)}, 
\enq
and the expansion of $f$ in terms of these polynomials reads, in the notation of \cite[section 3.5]{Dunkl01}
\beq
S_N(f)(x) = \sum_{n = 0}^N\sum_{|\vecn|,|\vecm| = n}s_\vecn \lb H_n^{-1}\rb_{\vecn\vecm}P_\vecm(x).
\enq
It will converge to the actual function $f$ for $N \rightarrow \infty$ if the set of polynomials is complete, whereas it may not if not. The expansion is also independent of the freedom there is in the choice of the orthogonal polynomials in equations (\ref{conditions}).  Writing the orthogonal polynomials in terms of a triangular transition matrix
\beq
P_\vecn(x) = \sum_{|\vecm|\le |\vecn|}C_{\vecn \vecm}x^\vecm
\enq
and taking $f$ as the score function $s(x,\alpha)$, it is simple to see that the independent bits of information of order $n$ are given by
\beq
\begin{split} \label{smulti}
s_n^2 &= \sum_{|\vecn|,|\vecm| = n} s_\vecn \lb H^{-1}_n \rb_{\vecn\vecm}s_\vecm \\
&=    \sum_{|\veci|,|\vecj| \le n} \lb C^TH_n^{-1}C\rb_{\veci \vecj} \frac{\partial m_\veci}{\partial \alpha}\frac{\partial m_\vecj}{\partial \alpha},
\end{split}
\enq
and the strict analog of equation (\ref{infn}) holds for each $N$,
\beq
\sum_{n= 1}^N s_n^2 = \sum_{|\veci|,|\vecj| = 1 }^N \frac{\partial m_\veci}{\partial \alpha}\lb \Sigma^{-1} \rb_{\veci\vecj} \frac{\partial m_\vecj}{\partial \alpha}.
\enq
\bibliographystyle{apj}
\bibliography{mybib}
\clearpage



\end{document}